\documentstyle[12pt,epsf]{article}

\newtheorem{guess}{Conjecture}
\newcommand{\half}{\frac{1}{2}}
\newcommand{\dt}{\triangle t}
\newcommand{\dz}{\triangle z}
\begin{document}
% the following command makes the equation numbers look like (2.3)
%\renewcommand{\theequation}{\thesection.\arabic{equation}}
\title{
 Drop Formation in a One-Dimensional Approximation \\
 of the Navier-Stokes Equation}

\author{
Jens Eggers\thanks{Present address: Universit\"at GH Essen, FB7, 
45117 Essen, Germany}
 \\Department of Mathematics and the James Franck
Institute \\ The University of Chicago, 5734 University Avenue \\
Chicago, Illinois  60637  U.S.A. 
\and
Todd F.\ Dupont \\Department of Computer Science \\ 
The University of Chicago, 1100 East 58th Street \\
Chicago, Illinois  60637  U.S.A.
      }
\maketitle
\begin{abstract}
We consider the viscous motion of a thin, axisymmetric column
of fluid with a free surface.  A one-dimensional equation of motion
for the velocity and the radius is derived from the Navier-Stokes
equation.  We compare with recent experiments on the breakup of a
liquid jet and on the bifurcation of a drop suspended from an orifice.
The equations form singularities as the fluid neck is pinching
off. The nature of the singularities is investigated in detail.

\end{abstract}
\vspace{3mm}
\section{Introduction}
A problem fundamental to the study of nonlinear partial differential
equations (PDE's) is the nature of their singularities.  Perhaps the
most famous (and unsolved) problem is the suspected blow-up of the
derivatives of the velocity field in the three-dimensional Euler
equation \cite{/Maj/}.  Shocks, i.e., discontinuities in the velocity, are
the type of singularities displayed by the one-dimensional inviscid
Burgers equation \cite{/Sm/}.

Still a different type of singularity has to be expected from
three-dimensional free surface flow, which we will consider
here. A similar study of a two-dimensional flow has been
conducted recently \cite{/CDG/}.  Surface tension will tend to make the
surface as small as possible by reducing the radius.   The classical
stability analysis of an infinite cylinder of fluid by Rayleigh
\cite{/Ray/} shows that the radius does not decrease uniformly:  Due to
the constraint of mass conservation the fastest growing mode is the
one with wavelength $\lambda \approx 9 r_0$, where $r_0$ is the radius
of the cylinder.  Consequently, the fluid cylinder will decay into
drops of roughly that size.

Once the radius becomes zero locally, i.e. the original column of fluid
separates, the description in terms of a radius function breaks down.
Hence the equations must develop a singularity at that point.
Although linear stability analysis gives a reasonable estimate of the
size of the droplets formed, it completely fails to predict the shape
of the surface once an appreciable deformation of the original
cylinder is reached \cite{/CR/}.  For example it does not
explain the fact that the cylinder does not break up uniformly.
Rather, regular size drops are, under most circumstances, followed by
much smaller ``satellite drops''.  Even higher order perturbation
theory \cite{/CR/,/CM/}  gives only a qualitative prediction of the
unequal drop sizes, but is not able to describe the shape of the fluid
anywhere close to pinch-off.  This is not very surprising, because
the characteristic time of the linear instability is close to the time
distance from the singularity, where expansions in the radius and the
velocity are bound to break down.

Therefore a complete treatment of the nonlinearities is needed.
The full Navier-Stokes equation with free boundary
conditions is extremely complicated, for both analytical and numerical studies.
The only simulations of axisymmetric drops we
are aware of were restricted to irrotational, inviscid flow \cite{/B/}.
But even with this restriction simulations close to the singularity
become extremely costly, since the neck region requires
high resolution.  A reduction of the problem to one dimension will
give huge savings in computer time, making closeups of the singularity
possible.

There already exist some one-dimensional equations for
axisymmetric free surface flow \cite{/Lee/,/Bo/}.  Lee \cite{/Lee/} only
considers the inviscid case.  Bogy's equations \cite{/Bo/} allow for
dissipation but are very complicated in structure and do not have a
clear connection with the original Navier-Stokes equation.  We will
therefore derive a set of one-dimensional equations by expanding the
radial variable in a Taylor series and keeping only the lowest order
terms of the Navier-Stokes equation.  Several invariances and conservation
laws of the Navier-Stokes equation are preserved.  This will be the
subject of the second section, along with a linear stability analysis.

Integrating the equations  near the singularity proves to be
very difficult, since the problem becomes very stiff due to the large
range of length scales in the problem.  We develop a fully implicit
centered difference method.  This scheme is then modified to treat the
convection term $vv_z$ by an upwinding technique which ensures
negative definiteness of the numerical dissipation.  The numerical
scheme is detailed in section three.

There is a fair amount of work applying one-dimensional equations to
the breakup of jets \cite{/Lee/,/Bo/}, liquid bridges \cite{/Me/} or
hanging drops \cite{/Cr/}. There is also work in this spirit on
films lining a cylindrical tube \cite{/JK/}.  
Yet a detailed comparison between
experiments and one-dimensional models within the nonlinear regime is
missing.  Therefore, we try to compare experimental drop profiles with
simulations close to the breakup point.  This is found in section four
for two recent experiments.

The first experiment \cite{/CM/} examines the decay of a free jet
of water,
the second observes how a hanging drop detaches after it is
adiabatically filled out of an orifice \cite{/PSS/}.  We also produce
an example with a high viscosity fluid.  Simulations
and experiments agree very well, giving ample support to the idea that
droplet breakup can be well described by one-dimensional equations.

In the next section we take a closer look at the pinch region.
We discuss a similarity theory for the nonviscous case and
explain its failure. All viscous solutions are determined by  
universal scaling functions close to the pinch point. 
The concluding section briefly discusses the approximation used in 
relation to other types of approximations, its higher order
versions, and the full Navier-Stokes or Euler equations.  
Finally, we indicate directions of future research.

\section{The equations of motion}

We start from the Navier-Stokes equation for an axisymmetric
column of fluid with kinematic viscosity $\nu$, density $\rho$, and surface tension $\gamma$.  In
cylindrical coordinates it reads \cite{/LL/}

\begin{equation}\label{(2.1)}
\partial_t v_r + v_r\partial_rv_r + v_z\partial_zv_r = -\partial_r p/\rho
\\
  + \nu(\partial^{2}_{r}v_r + \partial^{2}_{z}v_r + \partial_rv_r/r - v_r/r^2),
\end{equation}

\begin{equation}\label{(2.2)}
\partial_t v_z + v_r\partial_rv_z + v_z\partial_zv_z =
-\partial_z p/\rho \\
 + \nu(\partial^{2}_{r}v_z + \partial^{2}_zv_z + \partial_rv_z/r) - g,
\end{equation}
where $v_z$ is the velocity along the axis, $v_r$ the velocity in
the radial direction, and $p$ the pressure. The acceleration of 
gravity $g$ points in negative z-direction. 
The continuity equation reads

\begin{equation}\label{(2.3)}
\partial_rv_r + \partial_z v_z + v_r/r = 0 .
\end{equation}
The equations (\ref{(2.2)}) and (\ref{(2.3)}) hold for $0 \leq r < h(z,t)$.
The balance of normal forces gives

\begin{equation}\label{(2.4)}
{\bf n\; \sigma\; n} = \gamma(1/R_1 + 1/R_2) ,
\end{equation}
where ${\bf \sigma}$ is the stress tensor, ${\bf n}$ the
outward normal, and $R_1$ and $R_2$ are the principal radii of
curvature.  The tangential force balance is 

\begin{equation}\label{(2.5)}
{\bf n} \; {\bf \sigma} \; {\bf t} = 0 .
\end{equation}
Explicitly, this gives

\begin{equation}\label{(2.6)}
p/\rho - \frac{2\nu}{1+h'\, ^2}[\partial_rv_r + (\partial_zv_z)h'\, ^2 -
(\partial_rv_z + \partial_zv_r)h'] \\
= \frac{\gamma}{\rho}(1/R_1 + 1/R_2)|_{r=h}
\end{equation}
for the normal forces, and 

\begin{equation}\label{(2.7)}
\frac{\nu}{1+h'\, ^2}[2(\partial_rv_r)h' + (\partial_rv_z +
\partial_zv_r)(1 - h'\, ^2) - 2(\partial_zv_z)h'] = 0|_{r=h}
\end{equation}
for the tangential forces.  The prime refers to differentiation with
respect to $z$.  Finally, the surface has to move with the velocity
field at the boundary:

\begin{equation}\label{(2.8)}
\partial_t h + v_zh' = v_r|_{r=h}.
\end{equation}

Since we are going to look at thin columns of fluid relative to their
elongation, we expand in a Taylor series with respect to $r$.  By
symmetry we get

\begin{equation}\label{(2.9)}
v_z(z,r) = v_0 + v_2r^2 + \cdots ,
\end{equation}
and (\ref{(2.3)}) is satisfied by choosing $v_r$ to be

\begin{equation}\label{(2.10)}
v_r(z,r) = -v'_0r/2 - v'_2r^3/4 - \cdots  .
\end{equation}

The pressure is expanded in the same way:

\begin{equation}\label{(2.100)}
p(z,r) = p_0 + p_2r^2 + \cdots .
\end{equation}

We now insert (\ref{(2.9)})-(\ref{(2.100)}) into (\ref{(2.1)}), (\ref{(2.2)}), and
(\ref{(2.6)})-(\ref{(2.8)}) and solve the equations to lowest order in $r$.  In the
case of (\ref{(2.2)}) this gives

\begin{equation}\label{(2.11)}
\partial_tv_0 + v_0v'_0 = - p'_0/\rho + \nu(4v_2 + v''_0) - g.
\end{equation}
Equation (\ref{(2.1)}) is identically satisfied to lowest order. 

Remembering that
$h'$ is also of order $r$ we get from (\ref{(2.6)}) an expression for the
pressure $p_0$ in (\ref{(2.11)}):

\begin{equation}\label{(2.12)}
p_0/\rho + \nu v'_0 = \frac{\gamma}{\rho}(1/R_1 + 1/R_2)
\end{equation}
Similarly, (\ref{(2.7)}) gives an expression involving $v_2$:

\begin{equation}\label{(2.13)}
-v'_0h' + 2v_2h - v''_0h/2 - 2v'_0h' = 0.
\end{equation}
Equations (\ref{(2.12)}) and (\ref{(2.13)}) can be used to eliminate $p_0$ and $v_2$
from (\ref{(2.11)}) giving

\begin{equation}\label{(2.14)}
\partial_tv_0 = -v_0v_0' - \frac{\gamma}{\rho}(1/R_1 + 1/R_2)' +
3\nu(h^2v'_0)'/h^2 - g. 
\end{equation}
The surface condition (\ref{(2.8)}) says to lowest order 

\begin{equation}\label{(2.15)}
\partial_th = -v_0h' - v'_0h/2 .
\end{equation}
The formula for the mean curvature $\frac{1}{2}(1/R_1 + 1/R_2)$ of a
body of revolution is known from differential geometry \cite{/dC/}.
Thus, dropping the index on $v_0$ and denoting the surface tension
contribution of the pressure by $p$, we finally get

\begin{eqnarray}\label{(2.16)}
\partial_t v & = & - vv_z - p_z/\rho + 3\nu(h^2v_z)_z/h^2 - g , \\
p & = & \gamma\left[\frac{1}{h (1 + h^2_z)^{\half}} - \frac{h_{zz}}{(1+h^2_z)^\frac{3}{2}}\right] \nonumber
\end{eqnarray}
and

\begin{equation}\label{(2.17)}
\partial_th = -vh_z - v_z h/2 .
\end{equation}
Here the index $z$ refers to differentiation with respect to $z$.
When solving the set of equations (\ref{(2.16)}), (\ref{(2.17)}) for $z \in
[-\ell, \ell]$ we impose the boundary conditions

\begin{equation}\label{(2.18)}
h(\pm\ell, t) = h_\pm
\end{equation}
and

\begin{equation}\label{(2.19)}
v(\pm\ell, t) = v_\pm .
\end{equation}

The set of equations (\ref{(2.16)}) -- (\ref{(2.19)}) is going to 
concern us for the rest of this paper. We 
reiterate that the physical velocity field (\ref{(2.9)}), (\ref{(2.10)}) 
described by (\ref{(2.16)}), (\ref{(2.17)}) has both radial
and longitudinal components with a nontrivial r-dependence. 
The physical pressure (\ref{(2.100)}) also carries contributions
from the shear stress. This should be born in mind when we 
refer to $v$ and $p$ in (\ref{(2.16)}), (\ref{(2.17)}) 
as ``velocity'' and ``pressure''.

There are two important conservation laws for this simplified system.
First, mass conservation means

\begin{equation}\label{(2.20)}
\partial _tV = \pi h^2v|^{-\ell}_\ell ,
\end{equation}

\begin{equation}\label{(2.21)}
V = \pi \int^\ell_{-\ell} h^2dz .
\end{equation}
Second the sum of the kinetic energy
\begin{equation}\label{(2.22)}
E_{kin} = \frac{\pi}{2} \rho \int^\ell_{-\ell} h^2v^2dz ,
\end{equation}
and the potential energy

\begin{equation}\label{(2.24)}
E_{pot} = 2\pi\gamma \int^\ell_{-\ell} h \sqrt{1+h^2_z}\,dz
+ \pi\rho g \int^\ell_{-\ell} h^2 zdz
\end{equation}
obeys the balance equation
\begin{equation}\label{(2.25)}
\begin{array}{c}\partial_t(E_{kin} + E_{pot}) = \\
{\cal D} - \left.\pi\left( \frac{\rho}{2}
h^2v^3 - 
\frac{2\gamma\, hh_z\partial_th}{\sqrt{1+h^2_z}} +
ph^2v - 3\nu \rho vh^2v_z + \rho g h^2 v z
\right)\right|^\ell_{-\ell} . 
\end{array}
\end{equation}
So, apart from boundary terms the total energy changes with the rate of
energy dissipation

\begin{equation}\label{(2.26)}
{\cal D} = -3\pi\nu \rho \int^\ell_{-\ell} (hv_z)^2 dz .
\end{equation}
Since ${\cal D}$ is negative definite,
it follows that, without external driving (boundary terms in (\ref{(2.25)}))  ,
the total energy can only decrease. Note that the potential energy 
for the full equations is precisely (\ref{(2.24)}) , so that
the exact surfaces of static equilibrium are also equilibrium
surfaces of the model: they are states which minimize $E_{pot}$ \cite{/LL/}.
Famous examples are the equilibrium shapes of pendant drops \cite{/MW/}.
This was the reason for keeping lower order terms in the expression
for $p$: in a consistent expansion by orders of r the expression 
for $p$ simply would have been
\[p = \gamma (1/h - h_{zz}),\]
resulting in a different form of the potential energy.
We also note that ${\cal D}$ is {\em
not} negative definite for the viscous term as cited by Cram
\cite{/Cr/}.  His term $\nu v_{zz}$ may feed energy into the fluid,
which we found to prevent the system from reaching an equilibrium
state.

Although of limited applicability in practice, it is instructive to
repeat the stability analysis for a fluid cylinder in the case of our
model.  Assume a cylinder of radius $r_0$ receives a sinusoidal
perturbation of wavelength \mbox{$\lambda = 2\pi/k$}; then

\begin{eqnarray*}
r(z,t) & = & r_0[1 + \varepsilon(t)cos(kz)],\\
 v(z,t) & = & \varepsilon(t)v_0 sin(kz).
\end{eqnarray*}

Assuming $\varepsilon(t) = \varepsilon\, exp(\omega t)$, (\ref{(2.16)}) and
(\ref{(2.17)}) give to lowest order in $\varepsilon$

\[\omega v_0 = - \frac{\gamma}{\rho}(k/r_0 - r_0k^3) - 3\nu v_0k^2\]
and

\[ \omega = -v_0k/2 ,\]
respectively.  This leaves us with the dispersion relation

\begin{eqnarray}\label{(2.29)}
\omega^2 & = & \omega^2_0((r_0k)^2 - (r_0k)^4)/2 - 3\nu \omega k^2,\\
 \omega_0^2 & = & \gamma /r^3_0\rho .\nonumber
\end{eqnarray}
The solution of (\ref{(2.29)}) is

\begin{equation}\label{(2.30)}
\omega = \omega_0\left\{\sqrt{(kr_0)^2(1-(kr_0)^2)/2 + \frac{9}{4}\,
\frac{\ell_\nu}{r_0}(kr_0)^{4}\,} -
\frac{3}{2}(\frac{\ell_\nu}{r_0})^\half(kr_0)^2\right\} ,
\end{equation}
where 

\begin{equation}\label{(2.31)}
\ell_\nu = \nu^2\rho/\gamma 
\end {equation}
is a viscous length scale.  Both
the limits of zero viscosity,

\begin{equation}
\omega = \omega_0\sqrt{(kr_0)^2(1 - (kr_0)^2)/2}
\end{equation}
and high viscosity,

\begin{equation}
\omega = \frac{\gamma}{r_0\rho\nu}(1 - (kr_0)^2)/6 ,
\end{equation}
coincide with the exact result \cite{/Ch/} if an expansion to lowest
order in $kr_0$ is made.

Equation (\ref{(2.30)}) shows that there is an instability for long 
wavelengths, the stability boundary being $kr_0 = 1$ independent of $\nu$.
In the case of a random disturbance, however, the relevant quantity is
the {\em most unstable} or fastestest growing mode.  In the general
case this is 
\begin{equation}\label{(2.32)}
(k r_0)^2_{max}=\frac{1}{2 (1 + \sqrt{\frac{9}{2}\,\frac{\ell_\nu}{r_0}\;})}.
% old -- (k r_0)^2_{max}=\half\left(1+\sqrt{\frac{3}{2}\,\frac{\ell_\nu}{r_0}\;}\;\right).
\end{equation}
For $\nu=0$, the most unstable wavelength is therefore $\lambda_{max}
= 8.89\,r_0$ instead of the exact value of $9.01\,r_0$ \cite{/Ch/}.
In the limit of very high viscosity the infinite wavelength
perturbation becomes the most unstable one.

\section{Numerical procedure}

The numerical approximations were computed using a rather simple
finite difference scheme. The spatial mesh is highly nonuniform,
graded mesh; its refinement is based on the behavior of the computed
solution. The time-integration method is an adaptive fully implicit
$\theta$-weighted scheme.

Let the space mesh be
$$ z_1 < z_2 < \cdots < z_N$$
and adopt the following notation:
\begin{eqnarray*}
\dz_i & = &z_{i+1} - z_i,\\
z_{i+\half} & = & (z_i + z_{i+1})/2,\\
\dz_{i+\half} & = & z_{i+\frac{3}{2}} - z_{i-\half}.
\end{eqnarray*}
The meshes used were always constrained to satisfy
$$ \half \leq \frac{\dz_i}{\dz_{i+1}} \leq 2.$$

The solution at each time level is defined by two arrays, $\{h_i\}_{i=0}^N$
and $\{v_i\}_{i=1}^{N-1}$; the quantity $h_i$ is the
value of the approximate radius $h$ at the mesh point $z_i$ and the value $v_i$
gives the value of the approximate velocity $v$ at the point $z_{i+\half}$.
In describing the discrete equations for a particular time step it is 
convenient to let $dv_i$ and $dh_i$ denote the changes in $v_i$ and $h_i$,
respectively, that take place over the step.

Difference analogs of the $v$-equation, (\ref{(2.16)}),
were written corresponding to each point $z_{i+\half}$ and the difference
analogs of the $h$-equation, (\ref{(2.17)}), were written for each $z_i$.
The time derivative term was approximated by ${dv_i}/{\dt}$ or 
${dh_i}/{\dt}$, respectively.
The relation for $p$ was used to define it at each point $z_i$ in terms
of $h$ at $z_{i-1}$, $z_{i}$, and $z_{i+1}$, using centered differences
for the $h_z$ term and a second difference for $h_{zz}$. (Near the bottom
of a pendant drop this was changed; see the remarks just after
(\ref{(4.1)}).) This defines $p$ at each time level in terms of $h$ at
that level.

In setting up the difference equations that mimic (\ref{(2.16)}) and
(\ref{(2.17)}) the spatial terms (everything except the time derivative
terms) are evaluated using a weighted average of the current value and the
yet-to-be-computed value. 
These ``mid-step'' values can be written as $v_i + \theta dv_i$ and
$h_i + \theta dh_i$.
With $\theta = 0.5$ this gives
a second order correct in time difference equation, but we used $\theta$
slightly larger than $0.5$ (typically $\theta=0.55$). Using $\theta$ close
to one half gives a small first order truncation term (say 10\% of the
first-order backward difference equation). Taking $\theta > 0.5$ gives
smoother discrete solutions than $\theta = 0.5$.

The approximation of the $v v_z$ term at $z_{i+\half}$ is done as follows:
$$v v_z \;\cong\;  v_i (v_{i+1}-v_{i-1})/ \dz_{i+\half} + NVT $$
where $NVT$ is the numerical viscosity term that ``upwinds'' this
nonlinear convective term. The $NVT$ is structured so that it is 
an energy dissipation term of small size; the usual technique of
simply skewing the difference equation in the direction that 
the fluid is coming
from does not assure such a property. The $NVT$ term that we use is a
difference analog of
$$ \frac{-1}{h^2} (h^2 \tilde\nu (z) v_z )_z,$$
where $\tilde\nu (z) = \vartheta \; v \;\dz$ and $\vartheta$ is a nonnegative
parameter. Specifically,
\begin{eqnarray*}
\tilde{v}_{i+1} & = & \frac{ \dz_i v_{i+1} + \dz_{i+1} v_i }{ \dz_i + \dz_{i+1}}\\
h^2 \tilde\nu(z) v_z |_{i+1} & = & h^2_{i+1} (v_{i+1} - v_i)\; \vartheta \; \tilde{v}_{i+1}\\
NVT  & = & \frac{-1}{((h_{i+1}+h_i)/2)^2} \frac{h^2 \tilde\nu(z) v_z |_{i+1} - h^2 \tilde\nu(z) v_z |_{i}} {\dz_i}
\end{eqnarray*}
The rest of the $v$-equation formed as central differences. Note
associating $p$ with the $z_i$'s gives $p_z$ at the $z_{i+\half}$ points.
The viscosity term in (\ref{(2.16)}) is very similar to the NVT term;
the $\tilde\nu$-term is just the constant $\nu$.

The $h$-equation at $z_i$ has two spatial terms. The first, $v h_z$, is
approximated by 
\mbox{$\tilde{v}_i \, (h_{i+1} - h_{i-1})/(\dz_i + \dz_{i-1})$}. 
The second, $v_z \; h /2$, is approximated by
$$ \left(\frac{v_{i} - v_{i-1}}{\dz_i + \dz_{i-1}}\right) h_i.$$

In solving the nonlinear difference equations we use Newton's method. Many of
our simulations used only one Newton step per time step, starting from an
initial guess based on linear extrapolation from the previous two time levels.
It is quite easy, and reasonably efficient, to control the time step in such
a way that one step of Newton's method reduces the error to very close to
rounding error. It is worthwhile pointing out that even if the decision is
made to only use one step of Newton's method it is useful to code it in
general, since observing quadratic convergence of the iteration is a
good check on whether the linearization has been done correctly.

\section{Comparison with experiment}

The first experiment we consider studies the breakup of a liquid jet
\cite{/CM/}.  Water is pumped through a nozzle at high speed to form
a liquid column virtually unaffected by gravity.  A periodic
perturbation, whose amplitude and frequency can be controlled,
is applied to the jet as it leaves the nozzle,
The system is allowed to
reach a steady state, in which the jet at a sufficiently large
distance from the nozzle has completely broken up into droplets.
Photographs of this stationary configuration are taken.

We try to model the experiment as closely as possible, but since we
can only simulate up to the point of the first singularity
(due to limitations of our current program) we cannot
reach the stationary state.  Instead, we fix 
$h_+ = h_- \equiv r_0 \ll \ell$  and $v_+ = v_- \equiv V$, 
and over a period of 8 wavelengths smoothly turn on a small 
sinusoidal perturbation to $v_-$.

Thus the parameters of the simulation are the length of the jet $2\ell$,
its initial radius $r_0$, the fluid parameters $\gamma/\rho$ and
$\nu$, the speed of the jet $V$, and the amplitude $V_p$ 
and frequency $f_p$ of the
perturbation.  
We chose $r_0/2\ell = 0.004$, so the size of the drops is
very small compared with the jet length and the precise value
of this ratio is immaterial. $V_p$ was adjusted to 
make breakup times conform with experiment.
The remaining dimensionless parameters controlling the problem are
$\lambda / r_0$, $\ell_{\nu} / r_0$, and the Weber number 
$\beta^2 = \rho r_0 V^2/\gamma$.
Here $\lambda = V / f_p$ is the wavelength of the perturbation
and $\ell_{\nu}$ the viscous length (\ref{(2.31)}) .
Typical values for fluid parameters can be found in Table 1.
The jet experiments were done with water.   

For the jet, the linearized problem of Section 2 now takes place
in a semi-infinite geometry, where surface perturbations are prohibited 
to the left of the nozzle opening \cite{/KRT/},\cite{/LG/}.
However, for large Weber numbers (239 in the present experiment)
the growth of unstable modes is just the same as the temporal
growth of Section 2, translated into space via the jet velocity $V$.
Also, the parabolic velocity profile of the nozzle opening has 
relaxed into a plug profile in the relevant region of the jet \cite{/CM/},
so we are assuming a constant profile right from the opening. 

We follow the simulation up to the
first singularity.  The resulting profile is aligned with 
a picture of the experimental jet, to make the minima in front of 
the drop which is about to detach coincide.
In Figure 1 theoretical and experimental profiles are compared
for $\lambda/r_0 = 14.57$.
The case with the smallest
perturbation is shown \cite{/CM/}.
Allowing for some blur of the photographs, the agreement in the shape
of the drop about to form is quite nice. Note
that the breakup is taking place in a very asymmetric fashion (with
respect to the breakup point):  On the right side the profile is quite
steep forming a very much rounded drop; on the other side a flat neck
formed, which will eventually coalesce into a smaller satellite drop.

An even more direct comparison is possible with an experiment
investigating a dripping tap \cite{/PSS/}.  As long as the drop is
small, it will be suspended stably from the orifice.  By slowly
filling in more liquid, the drop goes through a series of stable
states, until eventually gravity overcomes surface tension and the
lower half of the drop falls.  Subsequently, a thin neck forms and the
lower part of the drop detaches.  The stability of the drop hanging in
equilibrium has been the subject of much study in itself \cite{/MW/}.
The length scale controlling this problem is the capillary length
\begin{equation}
\ell_c = (\gamma / \rho g)^{1/2} .
\end{equation}
For water, $\ell_c$ and $\ell_\nu$,
the viscous length, are separated by almost five orders of magnitude,
see Table 1.  Hence there is a wide range of physical phenomena to
explore between the onset of the linear instability and the breakup of
the drop.

We will not repeat the stability analysis for our one-dimensional
equations here, but concentrate on the breakup.  The only
dimensionless parameters in the problem are the ratios $\ell_c/r_0$ and
$\ell_\nu/r_0$, where $r_0$ is the radius of the orifice.  They were
made to coincide with the experimental values, the working fluid being
water.

There are some technical problems involved in simulating the moving
boundary at the lower end of the drop.  We avoid having to use a
movable grid by mapping the problem on the unit interval, $z/\ell = x
\in [0, 1]$ where $\ell$ is the length of the drop which is calculated
using 

\begin{equation}\label{(4.1)}
\ell(t) = \int^{t}_{t_0} v(1,s)ds.
\end{equation}

By definition, $v(1,s)$ is the velocity of the lower boundary.  Care
must also be taken to calculate the pressure at the endpoint where
$h_z$ becomes infinite.  For the values $x \in [0.9, 1]$ we calculate
$p$ by interpolating $h(x)$ with an {\em even} fourth-order
polynomial.  Then all the singularities in the mean curvature cancel.

Figure 2 shows a series of profiles taken at constant time intervals of
$0.4 (r_0^3\rho/\gamma)^{1/2}$.  In the experiment, this
would correspond to $6.6 ms$.  Given the very small time scale it
would be very costly to let fluid drip as slowly as is possible in
experiments.  To still let initial oscillations die out, the viscosity
is set to a very high value initially, and is then reduced to the
value of water well before the first instability. Fluid is injected 
at the orifice with speed $0.02\, \gamma /(\rho r^{1/2}_0)$.
To the profiles at constant time intervals we add a snapshot of the
drop as the width of the neck becomes $0.01\ r_0$.  We also
superimpose an experimental picture of the drop \cite{/PSS/}, 
taken at the point of breakup.

The very good agreement with simulations is especially impressive
since this was not to be expected from a simple one-dimensional
approximation.  In particular in the lower half of the drop the
assumption $h \ll \ell$ seems to fail, but one must remember that this
part of the drop is almost static in a moving frame of reference.  But
the static limit of the equations is retained exactly in the
approximation.  Note that although the linear instability of the
hanging drop is not investigated explicitly, it is also accurately
described by the model.  Namely, it determines the total volume of the
drop (upper and lower half combined) and influences the point of
breakoff.

Again, the breakoff occurs very asymmetrically, as was already
observed in the jet decay.  The asymmetry therefore does {\em not}
come from the action of gravity.  This is also confirmed by the
estimate of Peregrine et al. \cite{/PSS/}, who estimate that by the
time the neck is formed, straining forces due to surface tension
outweigh the straining forces due to gravity.

We conclude this section by reporting on a simulation of a fluid with
significantly higher viscosity.  With the radius of the orifice being
0.06 cm, we adjusted $\ell_c/r_0$ and $\ell_\nu/r_0$ to match the
parameter values for glycerol at $25^\circ C$, as given in Table 1.
The viscosity of glycerol is about 1,000 times higher than that of
water, leading to a very different type of dynamics.  Figure 3 shows the
neck being pulled into a long and thin thread. Its length is 
40 times the radius of the orifice at the point of rupture.
Qualitatively, this is consistent with linear stability
analysis: for high viscosity, the most unstable wavelength becomes
large, see equation (\ref{(2.32)}).  On the other hand, the radius $r_0$
of the thread becomes very small, so (\ref{(2.32)}) cannot account for
its length in any quantitative way.  The origin is clearly dynamical.
The break occurs at the upper end of the thread in the simulation
presented, but it may also happen close to the drop under slightly
different conditions.
Experiments with high viscosity fluids in the same geometry are in
in progress \cite{/SD/}.  

\section{Nature of Singularities}

We now look closer at the point where the fluid neck is pinching off.
As the radius goes to zero, pressure forces are expected to diverge,
and the small amount of fluid left in the neck region is pressed out
of it even faster.  Therefore, as $h_{min} \rightarrow 0$, where
$h_{min}$ is the minimum radius, the velocity and higher derivatives of
both $h$ and $v$ will probably become infinite at the point of
rupture.  This is the singularity or ``blow up'' we want to
investigate further.

Keller and Miksis \cite{/KM/} present a very interesting scaling
theory for the singularity in the nonviscous case.  There are two
important differences between our problem and theirs: Their Geometry is
two-dimensional rather than three-dimensional-axisymmetric, and they
study the time {\em after} the breakup.

The idea of their study may be described as follows:  Since $h$
becomes very small near the singularity and $v$ large, the pinch
region is separated in scale from the boundaries.  Therefore boundary
conditions become irrelevant and the flow is determined by
$\gamma/\rho$ alone.  If $\Delta t = t_s - t$ 
represents the time distance from the
singularity, the only available length scale is the combination
$(\gamma \Delta t^2/\rho)^{\frac{1}{3}}$.  Hence, introducing

\begin{equation}\label{(5.1)}
\overline{z} = (z - z_s)(\gamma \Delta t^2/\rho)^{- \frac{1}{3}} 
\end{equation}

where $z_s$ is the position of the pinch point, and 

\begin{equation}\label{(5.2)}
\overline{h} = h(\gamma \Delta t^2/\rho)^{- \frac{1}{3}}\, ,\ \ \ 
\overline{v} =
v(\gamma/\rho \Delta t)^{- \frac{1}{3}}\, 
\end{equation}
the problem can be written in terms of the similarity variables
$\overline{z}, \overline{h}$, and  $\overline{v}$ alone.  Once the
similarity equation is solved, the resulting profile determines the
evolution of the interface for all times up to the singularity.  Note
that $h \rightarrow 0$ and $v \rightarrow \infty$ as $\Delta t \rightarrow
0$ if $\overline h$ and $\overline v$ are assumed fixed, consistent
with the original assumptions.

We will see, however, that this similarity argument does not carry
through for the case of our equations, since the inviscid case
appears to develop singularities even before $h_{min} \rightarrow 0$ !  For a
consistent formulation up to the point of breakup we therefore need to
add at least a small amount of viscosity.
We are confident that the following conjecture, due to Constantin  \cite{/C/},
is true. It indicates that with
viscosity the singularity does not occur until $h_{min}$ goes
to zero.

\begin{guess}
For $\nu > 0$ and $t \in [0,t_0]$ such that $h(t) \geq h_0 > 0$ the
solutions of (\ref{(2.16)}), (\ref{(2.17)}) stay regular, i.e., $h$, $v$,
and all their derivatives remain bounded in $[-1, 1]$ for $t \in [0,
t_0]$ with bounds depending only on $\nu$ and $h_0$.
\end{guess}

To investigate the problem further,
we chose a cylinder of radius $r_0 = 0.01$ and length $2$ as an
initial condition.
At its ends $z = \pm 1$ the radius is kept fixed and the
velocity is set to zero.  Given a slight initial disturbance in the
velocity field, the cylinder collapses and forms a singularity after
about $20(r^3_0 \rho/\gamma)^{\half}$.  The viscous length is 
$\ell_{\nu} = 1.4 \cdot 10^{-5}$ .

The profile near the singularity comes out quite asymmetric, as
observed before. This is also true if one starts from
initial data almost symmetric around $z = 0$.  Two almost linear
pieces of different slope are joined smoothly by a round piece with a
radius of curvature comparable to the minimum radius, cf. Figure 4 . 
Hence both terms in the pressure in (\ref{(2.16)}) are of the same
order of magnitude at the minimum, while the first term dominates the
linear regime away from the minimum.  This means the pressure is higher
on the shallow side of the minimum (right hand side in Figure 4),
forcing fluid over to the steeper side (left hand side in Figure 4). 
As seen by comparing with the velocities in Figure 4, the
minimum is convected with the velocity of the fluid, so in the
frame of reference of the minimum fluid is expelled on either side.
At the same time this causes the left hand side to get even steeper.

If there is no mechanism curbing this process, the slope will
eventually get infinite.  All our simulations, conducted with
different initial radii and initial disturbances, show that
for the inviscid equations exactly this happens and $h_z$
goes to infinity {\em even before} $h_{min}$ goes to zero.  
Note that $\ell_{\nu}$ is still less than 2 \%
of the minimum height in Figure 4, yet the inviscid equations 
already would have blown up at the times shown. 
If $\nu$ is finite, on the other hand, $h_z$ turns out to be
uniformly bounded by a constant which gets larger as $\nu$ 
decreases. Hence for finite, but arbitrarily small $\nu$ 
blow-up only occurs for $h_{min} \rightarrow 0$. 

From the conjecture mentioned earlier we conclude that 
viscous solutions, up to some
finite minimum radius $h_0$, can be well approximated by finite
differences as long as the mesh is fine enough.   Intuitively, one
expects that the mesh size $\Delta z$ should be at least of the order
of $h_0$.  We checked convergence near the singularity by conducting a
series of runs with increasingly fine resolution.  To save on
computational effort, only the region around the singularity is highly
resolved, the grid getting coarser by factors of 2 towards the
outside.  We plotted $h_{min}$ and the maximum velocity $v_{max}$
versus the time difference from the singularity on logarithmic scales,
see Figure 5 . Lengths are shown in units of $\ell_{\nu}$ and
times in units of the viscous time scale
\begin{equation}\label{(5.3)}
t_{\nu} = \nu^3 \rho^2 / \gamma^2  
\end{equation}
The plots agree up to the length scale of the
coarser grid. We also monitored the highest derivatives in the
problem, i. e. $p_z$ and $v_{zz}$ . The dashed vertical line indicates 
up to which point they seemed well resolved. As can be seen in
Figure 5, problems with resolution occur when $h_{min}$ is of the 
order of $\Delta z$, indicated by the horizontal line. 
Since the numerical viscosity NVT as introduced in Section 3 
is approximately equal to $v \Delta z$, convergence for increasingly
fine grids also demonstrates that it does not introduce artificial
effects. For the finest resolution the numerical viscosity was 
less than a tenth of the physical viscosity in the center of the 
grid. From  
all this we feel confident that the plot of our best-resolved run in
Figure 5 gives a faithful description of the original equations up
to the point indicated.

Figure 5 indicates that $v_{max}$ goes to infinity as the singularity
$h_{min} \rightarrow 0$ is approached. All derivatives of the 
velocity as well as second or higher derivatives of the height 
are found to blow up even faster. This means their asymptotic value
increases faster than a negative power of $\Delta z$ as we increase 
the resolution. The maximum value of $h_z$, however, approaches a 
constant as $h_{min} \rightarrow 0$. This is an important 
self-consistency property of our equations: The solutions
never approach a situation where the surface parametrization is 
bound to break down.  
 
There are some regions where $h_{min}(t)$ is close to a 
power-law, but they never extend over more than two decades
in length scales. In $v_{max}$ there is even less an indication of 
power-law behavior. The decay of $h_{min}$ is always faster 
than the $t^{\frac{2}{3}}$ power-law predicted by
(\ref{(5.1)}), (\ref{(5.2)}). 

Considering also the profiles $h(z)$ and $v(z)$ directly,
we conclude that (\ref{(5.1)}), (\ref{(5.2)}) is clearly not
valid, even for $h_{min} \gg \ell_{\nu}$ . 
The reason may be that $(h_z)_{max}$ goes to infinity long before 
$h_{min}$ goes to zero, hence viscosity is important even 
on scales much larger than $\ell_{\nu}$. The nature of 
the singularity of our inviscid equations probably is specific 
to the approximation. For example, it could be that the full
Euler equations, instead of overturning, produce localized regions of 
high vorticity which our approximation cannot describe. 

The system described so far is determined by the four parameters
$r_0$, $\ell$, $\gamma/\rho$, and $\nu$ . If it has a solution $h(z,t)$
and $v(z,t)$, the system with parameters 
$ar_0$, $a\ell$, $(a^3/b^2)\gamma/\rho$, $(a^2/b) \nu$ will have
the scaled solution

\begin{eqnarray*}
h_{ab}(z,t) & = & a h(z/a,t/b),\\
v_{ab} & = & \frac{a}{b} v(z/a,t/b) .
\end{eqnarray*}

This is equivalent to saying that up to a rescaling of length
and time the solution is determined by {\em two} 
dimensionless ratios, $r_0/\ell$ and $\ell_{\nu}/\ell$, say.
By the argument presented at the beginning of this section,
one expects the solution near the singularity to be 
independent of the dimensions of the initial cylinder.
Hence {\em all} solutions, for $t \approx t_s$ and 
$z \approx z_s$, can be written in the universal form

\begin{eqnarray}\label{(5.4)}
h(z,t) & = & \ell_{\nu} h_s(\pm(z-z_s)/\ell_{\nu},(t_s-t)/t_{\nu}),\\
v(z,t) & = & \pm\frac{\ell_{\nu}}{t_{\nu}} 
v_s(\pm(z-z_s)/\ell_{\nu},(t_s-t)/t_{\nu}) . \nonumber
\end{eqnarray}

The $\pm$ signs in (\ref{(5.4)}) take care of the fact that
the solutions may have different parity, with fluid flowing 
from left to right or vice versa.

We tested (\ref{(5.4)}) by conducting simulations with 
different parameter values and calculating $h_s$ and $v_s$
from them. Namely, we increased $r_0$ by a factor of 10 
and also varied the viscosity. This causes the global
behavior of the solution to change dramatically, yet
on length and time scales comparable with $\ell_{\nu}$ and
$t_{\nu}$ or smaller (\ref{(5.4)}) is obeyed beautifully.
Figure 6 shows $h_s$ and $v_s$ as calculated from different
runs, all at $(t_s - t)/t_{\nu}$ = 1.97. Note that 
the reduced profiles still evolve in time, unlike the 
similarity solutions of (\ref{(5.1)}),(\ref{(5.2)}).

\section{Discussion}

The key to the success of the present investigation lies in
the construction of appropriate model equations to study
the motion of thin columns of fluid. First, our expansion 
method allows to take viscous body forces as well as
viscous boundary conditions into account. This makes it
distinct from methods where the average velocity over the
cross section is the dynamical variable, such as in the 
equations for shallow water waves \cite{/Per/}.
Precisely due to the inclusion of boundary conditions,
the viscous terms in our equations become purely dissipative.

Second, we take the exact curvature term (\ref{(2.16)}) 
into account. The importance of those higher order terms of the 
expansion for strong variations of $h$ was noticed before \cite{/JK/}.  
Figure 2, for example, beautifully demonstrates how the model takes 
equilibrium shapes into account. Also, regions of high slope
($h_z \approx 10$) at the top of the drop are very well represented.

Apart from experimental test, though, we do not see how to give
a priori estimates of the quality of approximation in the 
framework of our model alone. A possibility is to study the next 
order in the expansion. Apart from an equation of motion for
$h$, we now have {\em two} equations for the expansion coefficients of
the velocity field, $v_0(z,t)$ and $v_2(z,t)$. Those equations,
although readily written down, are considerably more complicated 
than (\ref{(2.16)}), (\ref{(2.17)}) and require new numerical
methods. Therefore, we consider it a study all of its own which
should be investigated separately.

Most importantly, our equations remain self-consistent right 
up to the point of rupture $h_{min} \rightarrow 0$. This
means no other singularity occurs before that point.
This is supported by our simulations over a wide range of 
viscosities and by preliminary mathematical analysis \cite{/C/}.
Specifically, there is no overturning of the profile.
The full equations of motion certainly would not form
singularities even in the case of overturning, but there does 
not seem to be experimental evidence for this to occur before 
breakup. (This excludes initial or boundary conditions 
with strong transversal velocity gradients, which ``force'' 
the flow to overturn, but which are not realizable in our 
equations in the first place.) Hence we see no reason to doubt
the applicability of our model even for small viscosities such
as in water. 

However, the inviscid version of our equations clearly is at
odds with experimental evidence, showing overturning on experimentally 
accessible timescales. This reflects the singular nature 
of the limit $\nu \rightarrow 0$. We hope this will shed some light 
on the nature of this limit in the Navier-Stokes equation 
and its relation to the Euler equation.

We plan to develop a
code with adaptable grid, which moves with the position of the minimum
and introduces new grid points when needed.  This code is expected to
be much more effective and to allow us to reach considerably higher
resolution. We hope this will allow us to explore the
asymptotic regime even more carefully.

Another expected benefit of the new code is to be able to go 
{\em beyond} the first singularity by introducing a new grid point
at the pinch. The equations will then be integrated from there
with new boundary conditions. This will allow us to investigate 
a new range of phenomena, like formation of satellite drops,
recoiling, etc.  

In conclusion, we have developed a one-dimensional equation for an
axisymmetric thread of fluid.  Computed profiles coincide nicely with
experiments.  The inviscid equations are inconsistent, leading to
singularities even before the breakup into drops. All solutions
with $\nu > 0$ are described by the universal scaling functions
$h_s$, $v_s$ sufficiently close to the singularity. 

\pagebreak

\begin{Large}
{\bf Acknowledgements}
\end{Large}
\vspace{3mm}

We are grateful to L. P. Kadanoff for getting us interested in
the problem, and to P. Constantin for discussions. D. Grier
helped tremendously with the image processing. 
J. E. thanks E. Becker for some early advice,
and the Deutsche Forschungsgemeinschaft for a fellowship.
J. E. is also supported by the ONR under grant No. N00014-90J-1194
and the NSF/DMR/MRL under grant No. 8819860.

\pagebreak

\pagebreak

\begin{center}
\bf TABLE 1
\end{center}

\begin{tabular}{|l|c|c|c|}
\hline
 & Water & Glycerol & Glycerol \\ & $20^\circ C$ & $20^\circ C$ &
$25^\circ C$ \\ \hline \hline $\nu\;\;[cm^2/sec]$ & $1.00 \cdot
10^{-2}$ & $11.8$ & $7.6$ \\ \hline $\gamma/\rho\;\;[cm^3/sec^2]$
fluid-air interface & $72.9$ & $50.3$ & 50.0
\\ \hline
$\ell_c=(\gamma/\rho g)^{1/2}\;\;[cm]$ & $0.273$ & $0.226$ & $0.226$ \\
\hline
$\ell_\nu=\rho\nu^2/\gamma\;\;[cm]$ & $1.38\cdot 10^{-6}$ & $2.79$ & $1.15$
\\ \hline
$t_\nu=\nu^3\rho^2/\gamma^2\;\;[sec]$ & $1.91\cdot 10^{-10}$ & $0.652
$ & $0.174$\\ \hline
\end{tabular}
\pagebreak

\begin{Large}
{\bf Table Captions}
\end{Large}
\vspace{3mm}

\begin{center}
\bf Table 1
\end{center}

This table contains the physical parameters for water at $20^{\circ}$C
and glycerol at $20^{\circ}$C and $25^{\circ}$C. The values are quoted
from \cite{/CRC/}.

The first line contains the kinematic viscosity $\nu$, the second the
surface tension divided by density $\gamma/\rho$. 
The remaining three lines contain
characteristic length and time scales, $g$ is the acceleration of
gravity.
\pagebreak

\begin{Large}
{\bf Figure Captions}
\end{Large}
\vspace{3mm}

\begin{center}
\bf Figure 1
\end{center}

Comparison between a decaying water jet \cite{/CM/} (upper) and our
simulation (lower). We processed the original image so as produce
a white background. 
The nozzle is to the left. The point where the
first drop detaches from the experimental jet has been aligned with
the corresponding point of our simulations. The horizontal scale has
been adjusted as well. The parameters are $\lambda/r_0 = 14.57$,
$\ell_{\nu}/r_0 = 6.54 \cdot 10^{-4}$, and $\beta^2=239$. The fluid 
parameters \cite{/CM/} differ slightly from the ones quoted in
Table 1, due to additives. 

\vspace{3mm}
\begin{center}
\bf Figure 2
\end{center}

Simulation of a drop of water suspended from a circular orifice of radius
$r_0 = 0.26cm$. This makes the parameters $l_c/r_0 = 0.992$ and 
$l_{\nu}/r_0 = 4.89 \cdot 10^{-6}$, compare Table 1.
The time distance between profiles is $0.4 (r_0^3\rho/\gamma)^{1/2}$,
starting from a point where the drop is already falling.
We also superimpose a profile at the snap-off and the 
corresponding experimental picture \cite{/PSS/}. There is no
adjustable parameter in the comparison. To enhance contrast,
we erased the background in the original 
photograph.

\vspace{3mm}
\begin{center}
\bf Figure 3
\end{center}

Same as Figure 2, but with the fluid being glycerol at $25^{\circ}$C and
$r_0 = 0.0625cm$. The parameters are now  $l_c/r_0 = 3.61$ and 
$l_{\nu}/r_0 = 18.3$ .
Note the long neck, which is the trademark of high viscosity fluids.

\pagebreak
\begin{center}
\bf Figure 4
\end{center}

A closeup view of simultaneous radius, velocity, and pressure  
profiles close to pinch-off. 
The time distance $\dt$ from the singularity are 
$\log_{10}(\dt) = -4.0$, $-4.1$, and $-4.2$, in units of 
$\gamma / \rho$ and $r_0$. The viscosity is $\nu = 0.0037$. 
The pressure
is higher on the right hand side of its peak, pushing fluid to the left.
The minimum of the radius $h$ moves with the fluid underneath it.

\vspace{3mm}
\begin{center}
\bf Figure 5
\end{center}

The minimum radius $h_{min}$, maximum (absolute) velocity $v_{max}$
and the maximum (absolute) slope $(h_z)_{max}$ as functions 
of the time distance from
singularity $\dt$, in units of the viscous scales $\ell_{\nu}$,
$t_{\nu}$, and $v_{\nu} =  \ell_{\nu} / t_{\nu}$. The axis
are logarithmic. The dashed vertical line indicates the point 
where $p_z$ is no longer fully resolved. This happens
when $h_{min}$ reaches the grid size. The $2/3$ - slope 
would be predicted by a nonviscous similarity theory.

\vspace{3mm}
\begin{center}
\bf Figure 6
\end{center}

The reduced profiles $h_s$ and $v_s$ as calculated from
different parameter values $(r_0,\ell,\gamma/\rho,\nu)$
via (\ref{(5.4)}). The solid line represents 
$(0.01,1,1,0.0037)$, the dotted line $(0.1,1,1,0.0037)$,
and the long-dashed line (0.01,1,1,0.0074).
The point of touch-down is shifted to zero in each
case, the units are the viscous scales. The dotted
lines had to be flipped over (- signs in (\ref{(5.4)}))
to correct for the difference in parity.

\pagebreak

%%%%%%%%%%%%%% Figure 1 %%%%%%%%%%%%%%
\newpage
\begin{figure}
  {\Huge Figure 1:}
  \begin{center}
    \leavevmode
    \epsfsize=0.9 \textwidth
    \epsffile{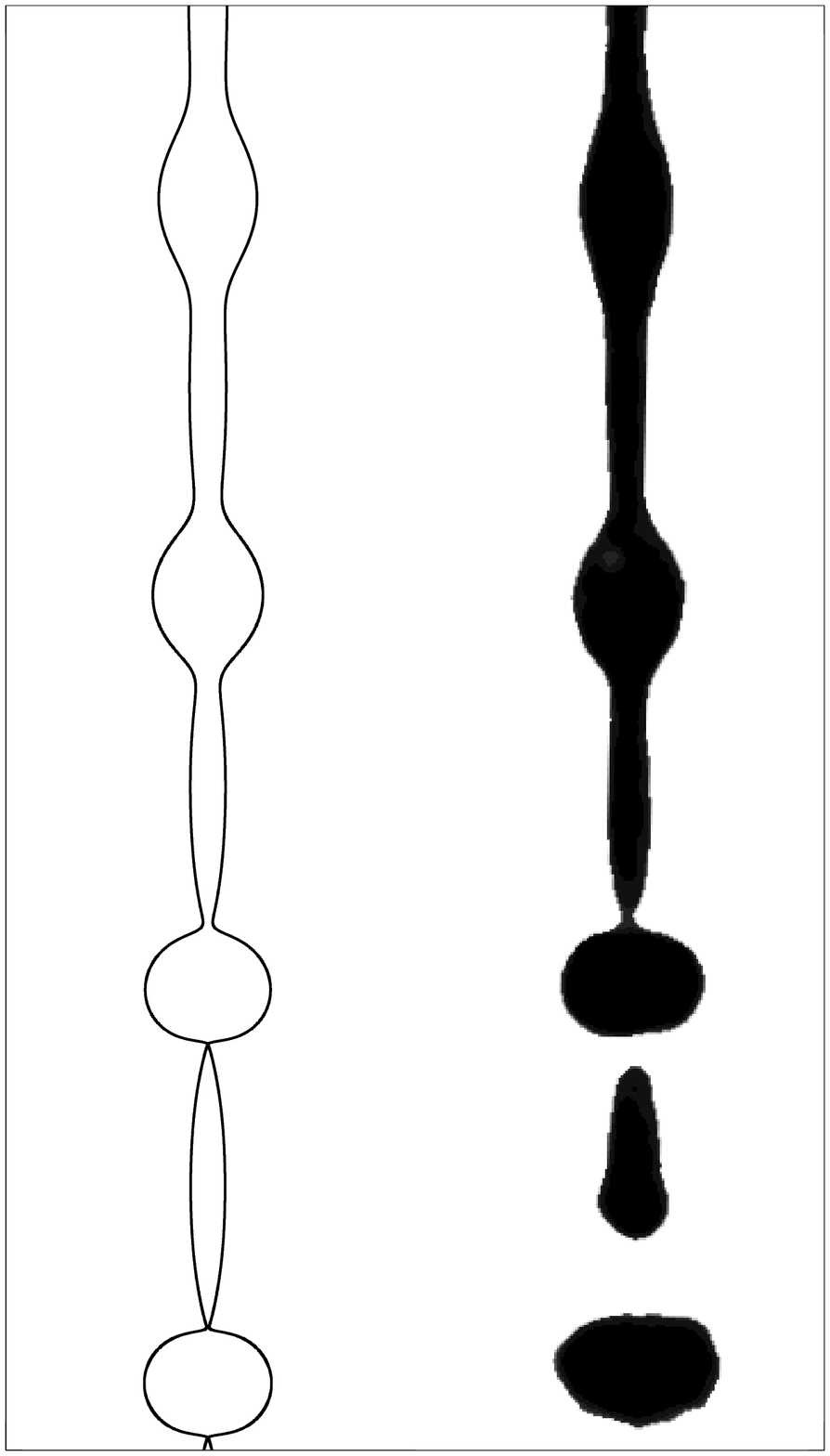}
  \end{center}
\end{figure}

%%%%%%%%%%%%%% Figure 2 %%%%%%%%%%%%%%
\newpage
\begin{figure}
  {\Huge Figure 2:}
  \begin{center}
    \leavevmode
    \epsfsize=0.9 \textwidth
    \epsffile{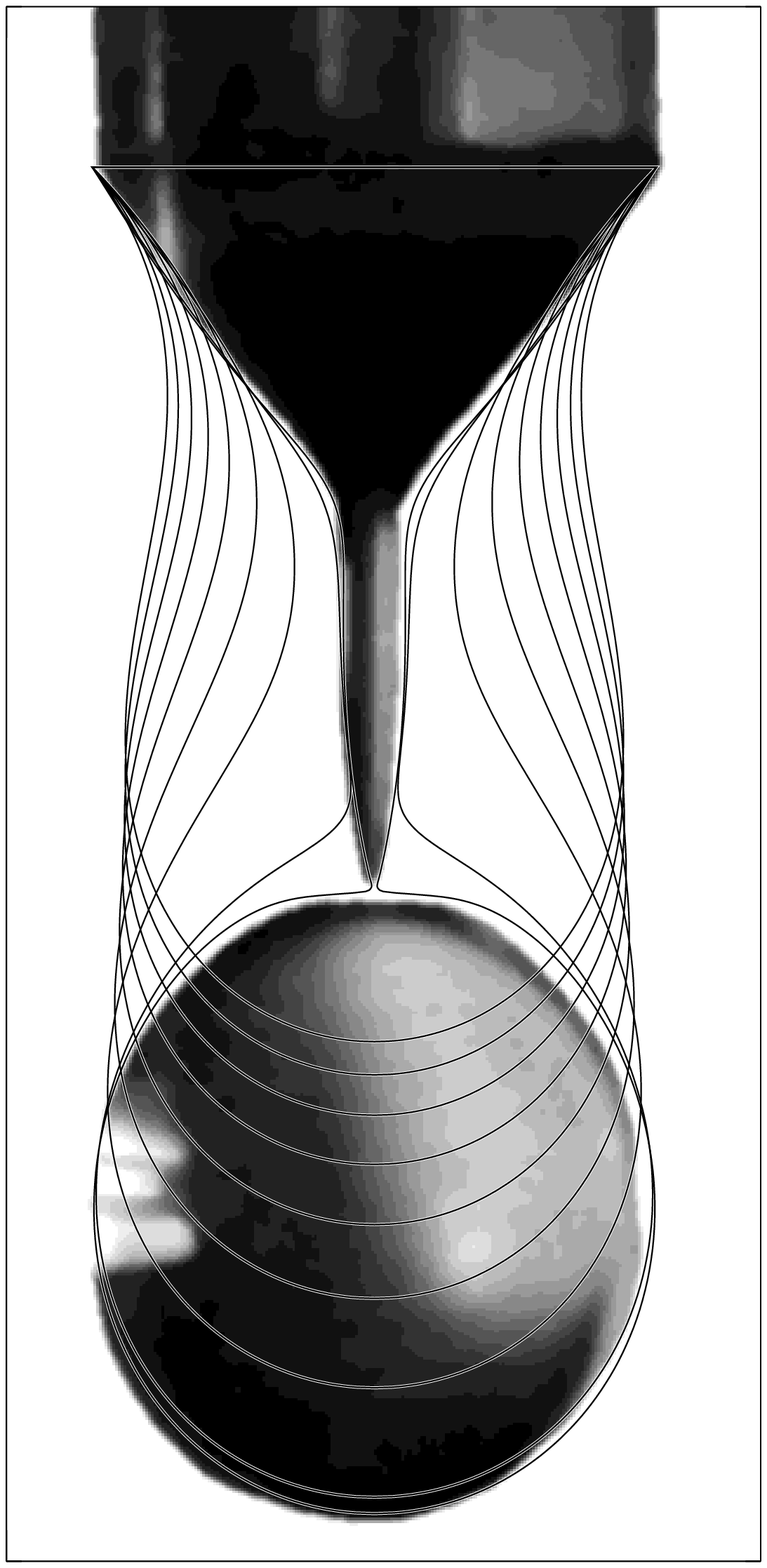}
  \end{center}
\end{figure}

%%%%%%%%%%%%%% Figure 3 %%%%%%%%%%%%%%
\newpage
\begin{figure}
  {\Huge Figure 3:}
  \begin{center}
    \leavevmode
    \epsfsize=0.9 \textwidth
    \epsffile{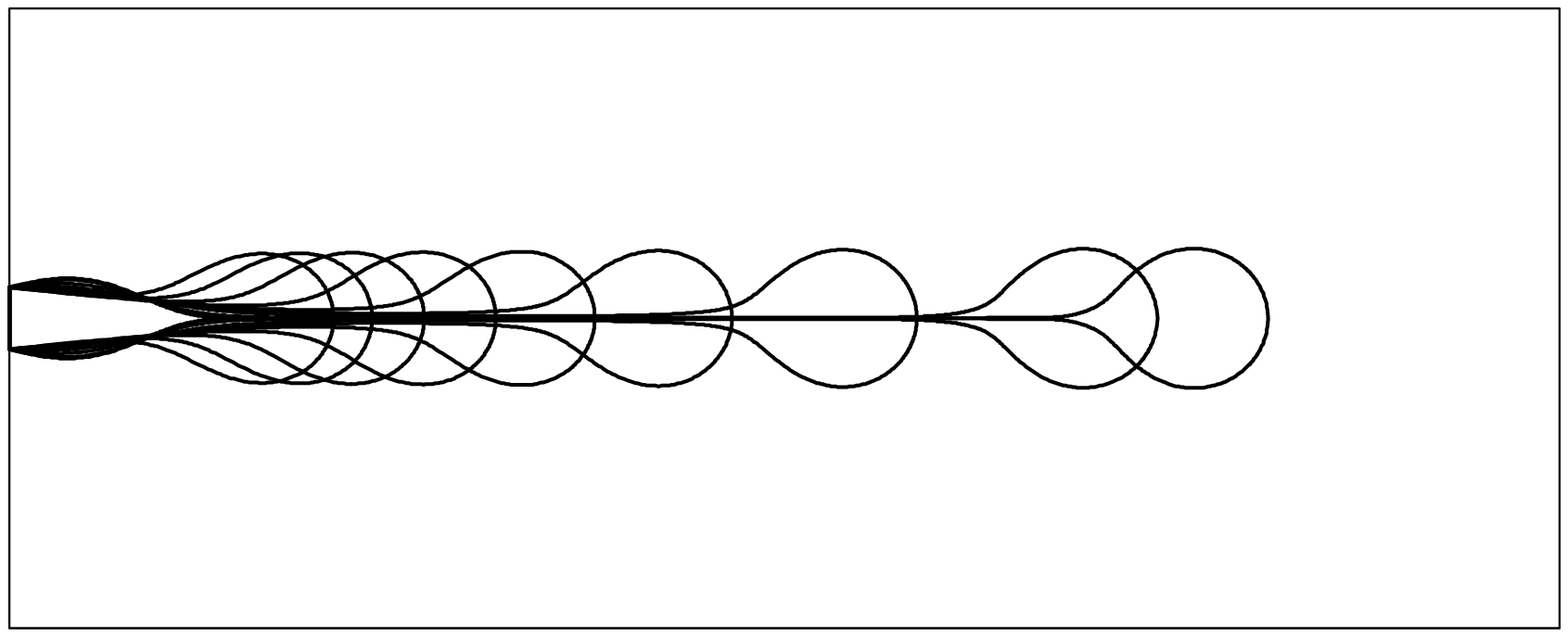}
  \end{center}
\end{figure}

%%%%%%%%%%%%%% Figure 4 %%%%%%%%%%%%%%
\newpage
\begin{figure}
  {\Huge Figure 4:}
  \begin{center}
    \leavevmode
    \epsfsize=0.9 \textwidth
    \epsffile{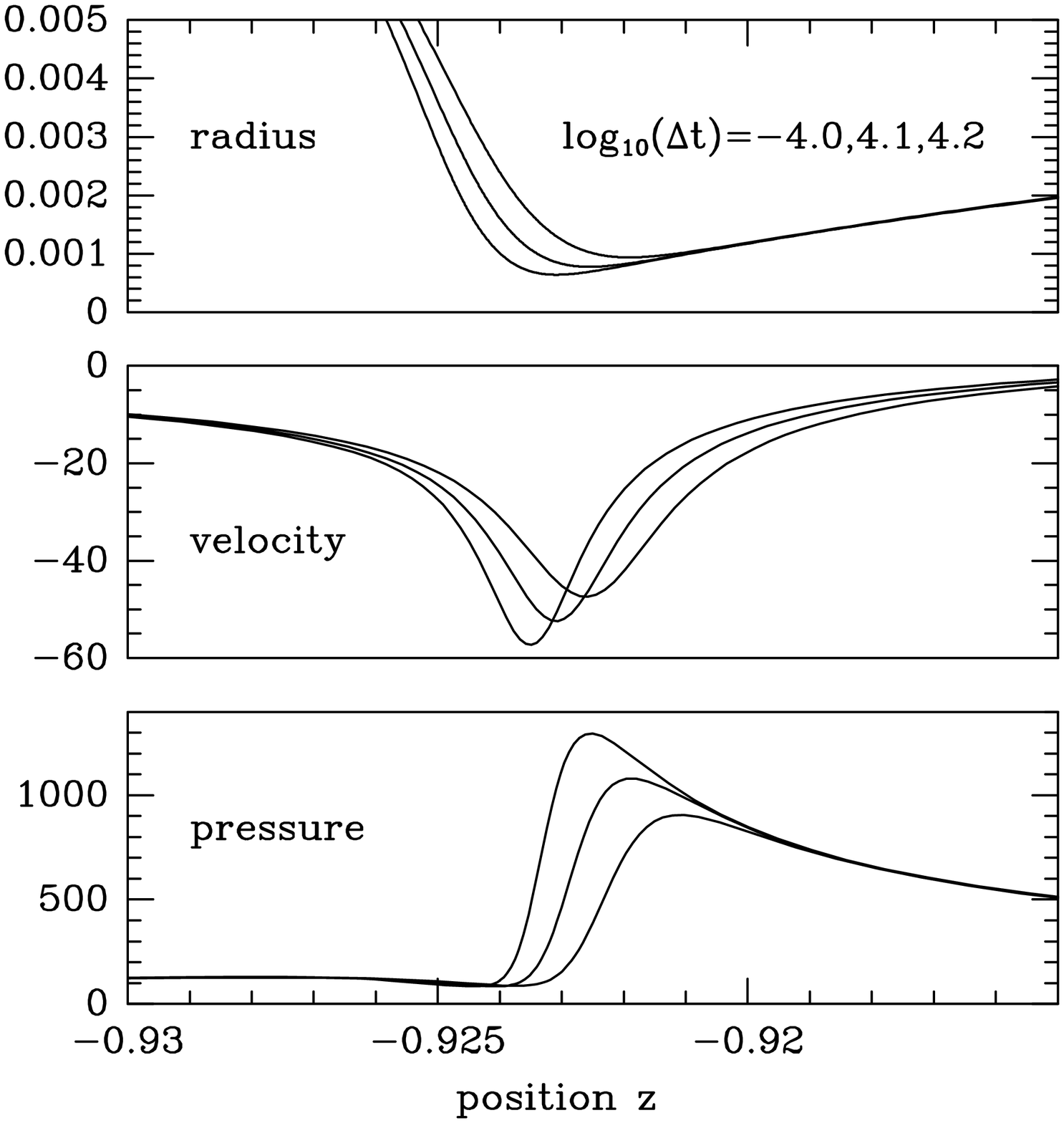}
  \end{center}
\end{figure}

%%%%%%%%%%%%%% Figure 5 %%%%%%%%%%%%%%
\newpage
\begin{figure}
  {\Huge Figure 5:}
  \begin{center}
    \leavevmode
    \epsfsize=0.9 \textwidth
    \epsffile{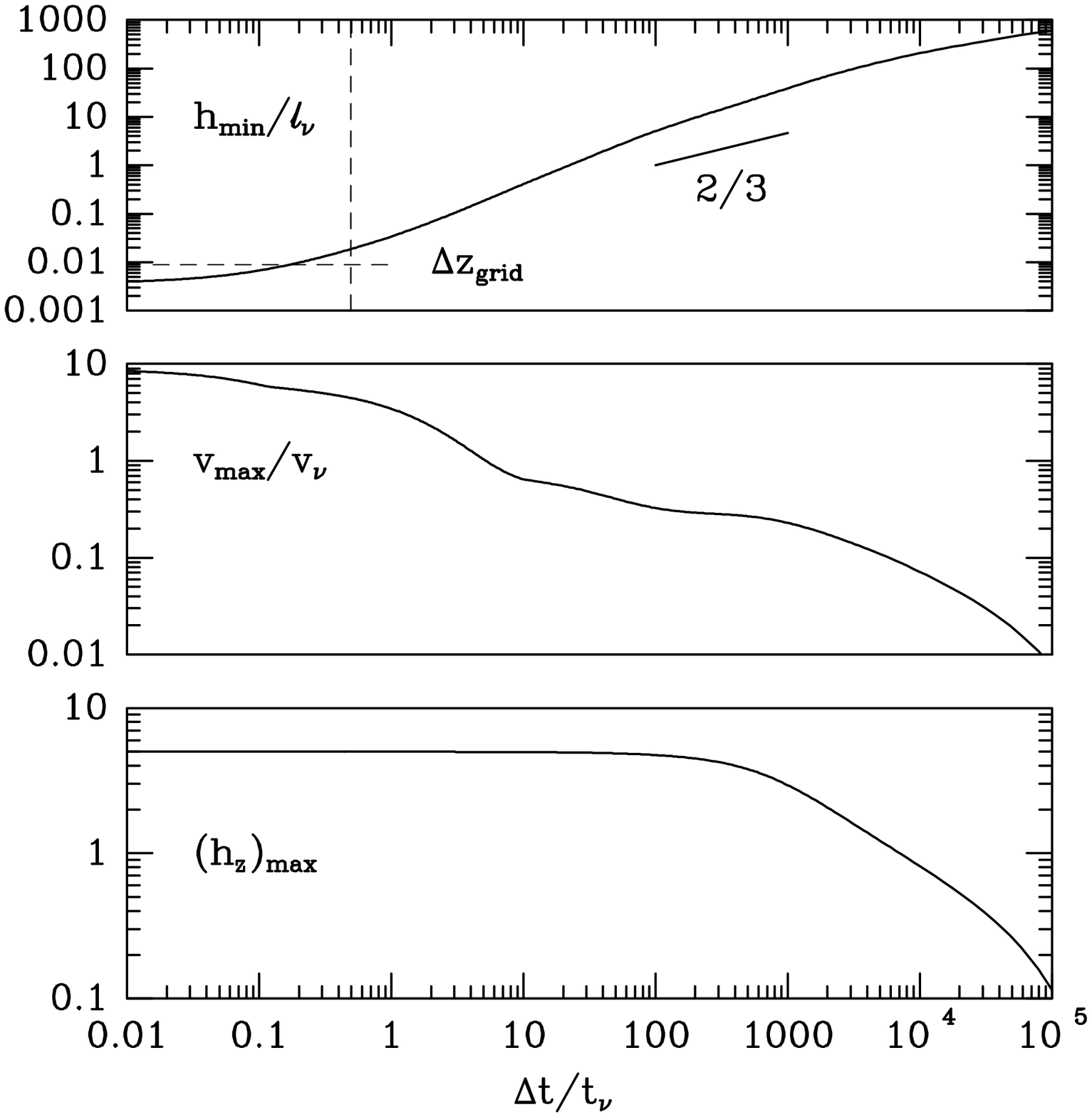}
  \end{center}
\end{figure}

%%%%%%%%%%%%%% Figure 6 %%%%%%%%%%%%%%
\newpage
\begin{figure}
  {\Huge Figure 6:}
  \begin{center}
    \leavevmode
    \epsfsize=0.9 \textwidth
    \epsffile{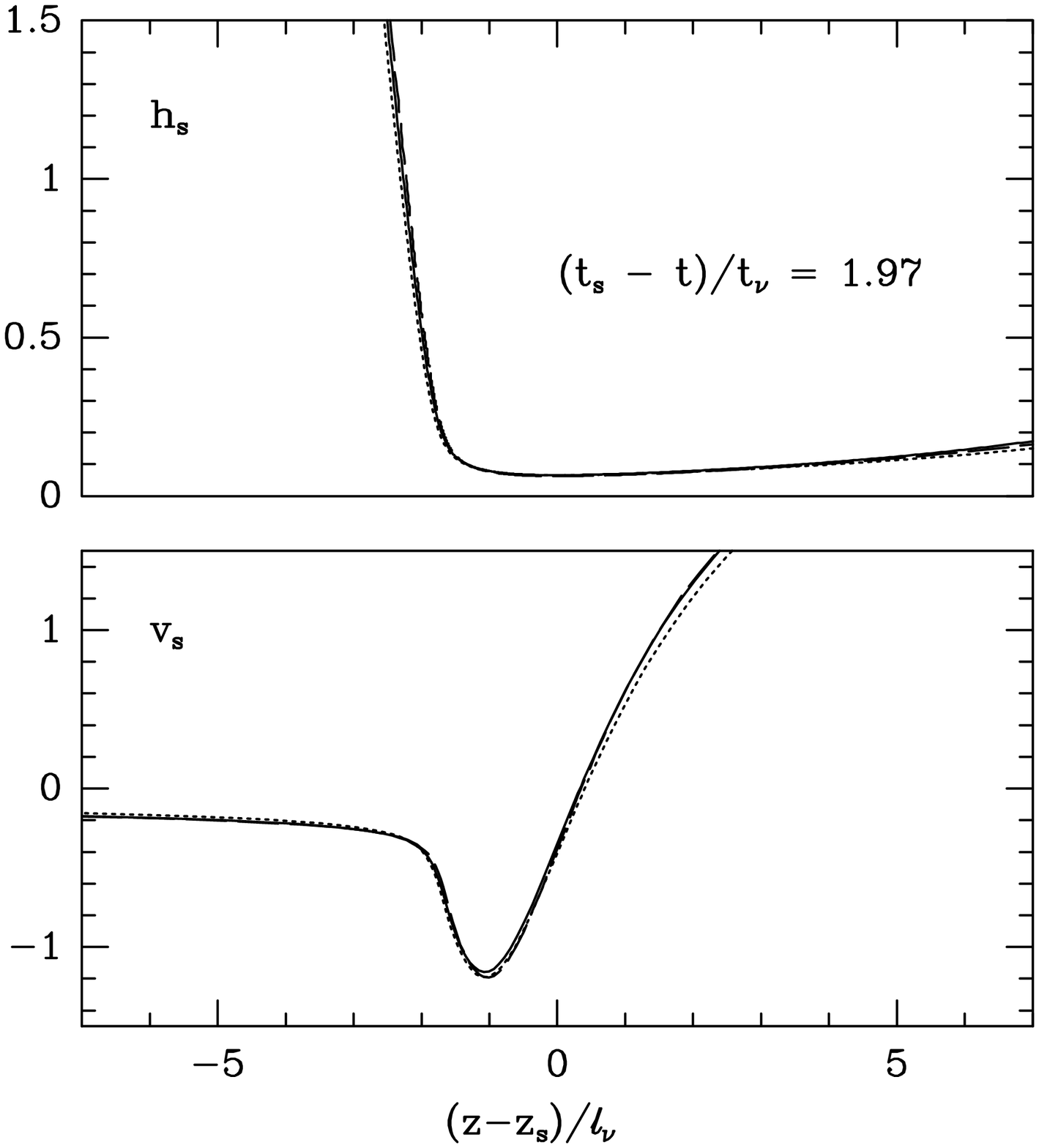}
  \end{center}
\end{figure}

\end{document}